\documentclass[showpacs,amsmath,amssymb,prl,superscriptaddress,twocolumn]{revtex4}
\usepackage[T1]{fontenc}
\usepackage{amssymb,amsmath}
\usepackage{times}
\usepackage{graphicx}
\usepackage{wasysym}
\usepackage{subfigure}
\usepackage[sort&compress]{natbib}
\usepackage{bm}
\usepackage{overpic}
\usepackage{color}

\begin{document}

\title{Repulsive Casimir Force in Chiral Metamaterials}

\author{R.~Zhao}
\affiliation{Ames Laboratory and Dept.~of Phys.~and Astronomy,
             Iowa State University, Ames, Iowa 50011, U.S.A.}
\affiliation{Applied Optics Beijing Area Major Laboratory, Department of Physics,
Beijing Normal University, Beijing 100875, China}

\author{J.~Zhou}
\affiliation{Ames Laboratory and Dept.~of Phys.~and Astronomy,
             Iowa State University, Ames, Iowa 50011, U.S.A.}

\author{Th.~Koschny}
\affiliation{Ames Laboratory and Dept.~of Phys.~and Astronomy,
             Iowa State University, Ames, Iowa 50011, U.S.A.}
\affiliation{Institute of Electronic Structure and Laser, FORTH,
             and Department of Materials Science and Technology, University of Crete,71110 Heraklion, Crete, Greece}

\author{E.~N.~Economou}
\affiliation{Institute of Electronic Structure and Laser, FORTH,
             and Department of Materials Science and Technology, University of Crete,71110 Heraklion, Crete, Greece}
\affiliation{Department of Computational and Data Sciences, George Mason University, Fairfax, Virginia 22030, USA}

\author{C.~M.~Soukoulis}
\affiliation{Ames Laboratory and Dept.~of Phys.~and Astronomy,
             Iowa State University, Ames, Iowa 50011, U.S.A.}
\affiliation{Institute of Electronic Structure and Laser, FORTH,
             and Department of Materials Science and Technology, University of Crete,71110 Heraklion, Crete, Greece}
\date{\today}


\begin{abstract}
We demonstrate theoretically that one can obtain repulsive Casimir
forces and stable nanolevitations by using chiral metamaterials. By
extending the Lifshitz theory to treat chiral metamaterials, we find
that a repulsive force and a minimum of the interaction energy exist
for strong chirality, under realistic frequency dependencies and
correct limiting values (for zero and infinite frequencies) of the
permittivity, permeability, and chiral coefficients.
\end{abstract}



\maketitle

Following the original Casimir paper \cite{Casimir} for the
attraction of two media, 1 and 2 occupying half spaces, $z<0$ and
$z>d$, respectively, and such that the electromagnetic fields are
confined exclusively in the vacuum region between them, Lifshitz
\cite{Lifshitz} generalized the calculation of this force to the
case that these two media are characterized by frequency-dependent
dielectric functions $\epsilon_1(\omega)$ and $\epsilon_2(\omega)$.
Subsequently, there was further generalization to general
bi-anisotropic media \cite{Barash}. The formula for the force or
the interaction energy per unit area can be expressed in terms of
the reflection amplitudes, $r_j^{ab}$ ($j=1,2$) \cite{Lambrecht}, at the vacuum/medium
$j$ interface, giving the ratio of the reflected EM wave of
polarization \textit{a} by the incoming wave of polarization
\textit{b}. Each \textit{a} and \textit{b} stands for either
electric (TM or p) or magnetic (TE or s) waves. The frequency
integration is completed along the imaginary axis by setting
$\omega=i\xi$. The formula for the interaction energy per unit
area becomes \cite{another}
\begin{equation}\label{energy}
\frac{E(d)}{A}=\frac{\hbar}{2\pi}\int_0^\infty d\xi\int\frac{d^2\mathbf{k}_\parallel}{(2\pi)^2}\ln\det \mathbf{D},
\end{equation}
where $\mathbf{D}=1-\mathbf{R}_1\cdot\mathbf{R}_2e^{-2Kd},K=\sqrt{\mathbf{k}_\parallel^2+\xi^2/c^2}$, and
\begin{equation}\label{matrix}
\mathbf{R}_j=\left\lvert\begin{array}{cc}
r_j^\textrm{ss}&r_j^\textrm{sp}\\r_j^\textrm{ps}&r_j^\textrm{pp}
\end{array}\right\rvert.
\end{equation}
For isotropic media, the off-diagonal terms in Eq. (\ref{matrix}) vanish and
\begin{equation}\label{element1}
r_j^\textrm{ss}=\frac{\mu_jK-K_j}{\mu_jK+K_j},~r_j^\textrm{pp}=\frac{\epsilon_jK-K_j}{\epsilon_jK+K_j};~j=1,2,
\end{equation}
where $K_j=\sqrt{\mathbf{k}_\parallel^2+\epsilon_j\mu_j\xi^2/c^2}$ and $\mu_j$ is the permeability of medium $j$.

In most cases the resulting Casimir force between the two media
separated by a vacuum region is attractive. There is increased
interest recently \cite{repulsive1,repulsive2,repulsive3,repulsive4}
in determining whether there is a combination of media 1 and 2
capable of producing a repulsive force. There have been mainly three
mechanisms to obtain repulsion for the Casimir force: (1)
Dzyaloshinskii's Casimir repulsion \cite{repulsive1}: Immersing the
interacting plates of $\epsilon_1$ and $\epsilon_2$ in a fluid of
$\epsilon_3$ and, moreover, satisfying the condition
$\epsilon_1(i\xi)<\epsilon_3(i\xi)<\epsilon_2(i\xi)$; (2) Boyer's
Casimir repulsion \cite{repulsive2}: Based on an asymmetric setup of
mainly (purely) nonmagnetic/vacuum/mainly (purely) magnetic; (3)
Leonhardt's Casimir repulsion \cite{repulsive3}: Employing a perfect
lens sandwiched between the interacting plates. The possibility for
a transition from an attractive to a repulsive force as the distance
\textit{d} decreases (corresponding to a minimum of the
interaction energy) leads to nanolevitations and opens up many
opportunities for application, e.g., almost frictionless operation
of nanomotors. Even through Capasso's group experimentally
realized the repulsion, based on the theoretical prediction of
Dzyaloshinskii et al. \cite{repulsive1}, this kind of system
still has friction because of the existence of the liquid.
Leonhardt's Casimir repulsion needs a perfect lens with
simultaneously negative dielectric permittivity and magnetic
permeability, which are extremely difficult to obtain at optical wavelengths.
Finally, Boyer's Casimir repulsion proposal faces the essential
obstacle that such nontrivial magnetic materials in the optical
regime do not exist in nature, and, therefore, it relies on the
nontrivial possibility of developing new artificial negative index
metamaterials (NIMs).

In this letter, we examined  \textit{realistic} non-chiral
metamaterials and we concluded they do not give a repulsive
Casimir force. However, we found that chiral metamaterials are
excellent candidates to realize the repulsive Casimir force. The
existence of a repulsive Casimir force depends upon the strength of
the chirality. We present analytical arguments that strong chirality
gives a repulsive force, supported by numerical calculations.

Negative index metamaterials  \cite{NIMs}, because of their
resonance magnetic response, offer more flexibility and, hence, more
promise for achieving a repulsive Casimir force, based on Boyer's
prediction. Indeed, in recent papers, Rosa et al. \cite{Rosa} found
a repulsive force in a range of values of $d$ for a combination of a mainly nonmagnetic
Drude-modeled silver and a magnetic NIM. This result was obtained \cite{Rosa} through the employment of a Lorentz type of magnetic permeability of the form
$\mu(\omega)=1-\Omega^2/(\omega^2-\omega_m^2+i\gamma\omega)$. This
form provides the opportunity to use an $\Omega$
large enough as to satisfy the condition $\mu(i\xi)>\epsilon(i\xi)$ and obtain
thus Boyer's Casimir repulsion. For the reasons stated below we consider a Lorentz type frequency dependence of $\mu(\omega)$ unphysical. Instead we employed the following realistic expression for $\mu(\omega)$: 
\begin{equation}\label{mu}
\mu(\omega)=1+\alpha-\frac{A\omega^2}{\omega^2-\omega_m^2+i\gamma_m\omega},
\end{equation}
where $|\alpha|$ is usually much smaller than one and $A=\alpha$ in order to satisfy the physical requirement that $\mu(\omega)\to1$ as $\omega\to\infty$. It must be stressed that the realistic expression (\ref{mu}), although almost identical to the Lorentz form for $\omega$ around the resonant value $\omega_m$, produces radically different results than the Lorentz one as far as the Casimir attraction is concerned. As the authors of Ref. \cite{Rosa} have found out (and we have confirmed), expression (\ref{mu}) (with $\alpha=0$ and $A\neq0$) combined with the form of Eq. (\ref{epsilon}) below for $\epsilon(\omega)$ does not produce repulsion. This is also true for the realistic case of $\alpha=A$.

The $\omega^2$ dependence of the numerator of the resonance term
follows from the equivalent circuit approach \cite{Pendry} and from the Maxwell's equations in the low frequency regime as stated in Ref. \cite{Rosa}. It is
confirmed by the retrieval procedure in actual SRR based and fishnet metamaterials. Of
course, it is possible to have more than one resonance term in Eq.
(\ref{mu}), but their coefficients must satisfy the relation $\alpha-\sum_i A_i=0$
to obtain the correct limiting value of $\mu(\infty)=1$. Besides cases
having $\alpha=A$, we also examine the case  $\alpha=0$ and
$A\neq0$ (which produces the incorrect limiting behavior,
$\mu(\omega)=1-A$ as $\omega\to\infty$). The reason for this
unphysical choice is to determine the role of the $\omega=\infty$
value of $\mu(\omega)$. The most general form of the frequency-dependence of the dielectric function is the sum of the Drude term and
several Lorentz-type resonance terms. If only one resonance term is
kept, we have
\begin{equation}\label{epsilon}
\epsilon(\omega)=1-\frac{\omega_{pl}^2}{\omega^2+i\gamma_{pl}\omega}-\frac{\omega_e^2}{\omega^2-\omega_R^2+i\gamma_R\omega}.
\end{equation}

\begin{figure}[htb!]
 \centering{\includegraphics[angle=0, width=8.cm]{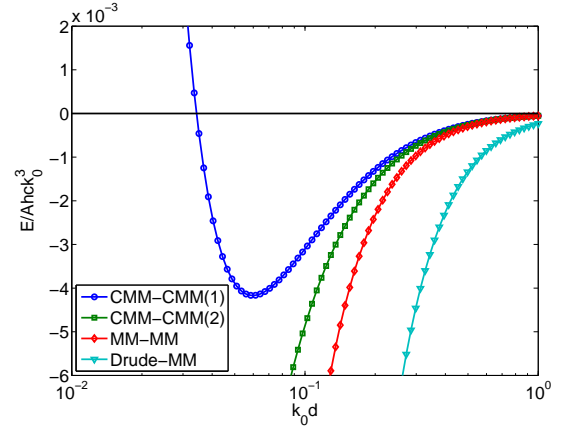}}
 \caption{(Color online) Casimir interaction energy per unit area $E/A$ (in
units of $hck_0^3$) versus $k_0d$; $k_0=\omega_R/c$. The triangle
curve corresponds to $\alpha=A=0.001,~\kappa=0$ (no
chirality),$~\omega_m=\omega_R,~\gamma_m=\gamma_R=0.05\omega_R,~\omega_{pl}=0,~\omega_e=\omega_R$
for material 1, while
$\alpha=A=0,~\omega_{pl}=10\omega_R,~\gamma_{pl}=0.05\omega_{pl},~\omega_e=0$
for material 2. The diamond curve is the case with
$\alpha=A=0.001,~\kappa=0$,
$~\omega_m=\omega_R,~\gamma_m=\gamma_R=0.05\omega_R,~\omega_{pl}=0,~\omega_e=\omega_R$.
The squares curve is the case with
$\alpha=A=0.001,~\omega_{\kappa1}=\omega_{\kappa2}=0.6\omega_R,~\omega_m=\omega_{\kappa
R}=\omega_R,~\gamma_m=\gamma_\kappa=\gamma_R=0.05\omega_R,~\omega_{pl}=0,~\omega_e=\omega_R$.
Finally, the circle curve shows repulsion for $k_0d<0.0586$ and a stable equilibrium point at
$k_0d=0.0586$; the parameters are the same as for the
square curve except for $\omega_{\kappa1}=\omega_{\kappa2}=0.7\omega_R$.}
 \label{Fig1}
\end{figure}

We have calculated the Casimir force using for material 1 and
material 2, $\epsilon_1, \mu_1$ and $\epsilon_2, \mu_2$, as in Eqs.
(\ref{mu}) and (\ref{epsilon}) with several values of $A,~\omega_m,~\omega_{pl}^2,~\omega_e^2$ (including
$\omega_{pl}^2=0,~\omega_e^2\ne0$, and
$\omega_{pl}^2\ne0,~\omega_e^2=0$). Among these values, we included
realistic values as they were obtained by our retrieval approach in
various fabricated and/or simulated NIMs. The Casimir force turned
out to be attractive in all cases we calculated. See the triangle
and diamond curves in Fig. \ref{Fig1}.

Recently, a lot of experimental work on chiral metamaterials (CMMs)
fabricated by planar technologies have been published \cite{chiral}.
For such artificial materials, the constitutive equations have the
form
\begin{equation}\label{constitutive}
\left(\begin{array}{cc}\mathbf{D}\\\mathbf{B}\end{array}\right)=\left(\begin{array}{cc}
\epsilon_0\epsilon&i\kappa/c_0\\-i\kappa/c_0&\mu_0\mu
\end{array}\right)\left(\begin{array}{cc}\mathbf{E}\\\mathbf{H}\end{array}\right)
\end{equation}
where the coefficient $\kappa$ has the following frequency
dependence for the chiral metamaterials \cite{kappa}:
\begin{equation}\label{kappa}
\kappa(\omega)=\frac{\omega_\kappa\omega}{\omega^2-\omega_{{\kappa}R}^2+i\gamma_{\kappa}\omega},
\end{equation}
which is the same as Condon model for homogeneous chiral molecular media \cite{Condon}.

For such CMMs, the reflection elements can be expressed as follows,
assuming the electromagnetic wave is from vacuum to chiral
metamaterials \cite{Lakhtakia},
\begin{subequations}\label{elements}
\begin{align}
r_j^\textrm{ss}&=\frac{-\Gamma_-(\chi_++\chi_-)-(\chi_+\chi_--1)}{\Gamma_+(\chi_++\chi_-)+(\chi_+\chi_-+1)},\\
r_j^\textrm{pp}&=\frac{\Gamma_-(\chi_++\chi_-)-(\chi_+\chi_--1)}{\Gamma_+(\chi_++\chi_-)+(\chi_+\chi_-+1)},\\
r_j^\textrm{sp}&=\frac{i(\chi_+-\chi_-)}{\Gamma_+(\chi_++\chi_-)+(\chi_+\chi_-+1)},\\
r_j^\textrm{ps}&=-r_j^\textrm{sp},
\end{align}
\end{subequations}
and
\[\chi_\pm=\frac{K_\pm}{n_\pm K},~~~\Gamma_\pm=\frac{\eta^2_0\pm\eta^2_j}{2\eta_0\eta_j},\]
where, $K_\pm=\sqrt{\mathbf{k}_\parallel^2+n_\pm^2\xi^2/c^2}$,
$n_\pm(i\xi)=\sqrt{\epsilon_j(i\xi)\mu_j(i\xi)}\pm \kappa_j(i\xi)$, $\eta_0=\sqrt{\mu_0/\epsilon_0}$,
$\eta_j=\sqrt{\mu_0\mu_j(i\xi)/\epsilon_0\epsilon_j(i\xi)}$,
$\epsilon_j(i\xi)$ and
$\mu_j(i\xi)$ are the relative permittivity and
permeability of the plate $j$, respectively, and $\kappa_j(i\xi)$ is the chirality coefficient. Although $n_\pm$ are complex, the reflection elements, r's, are still purely real because $\chi_+=\chi_-^\ast$.

\begin{figure}[htb!]
   \subfigure{\label{Fig2a}\includegraphics[width=0.23\textwidth]{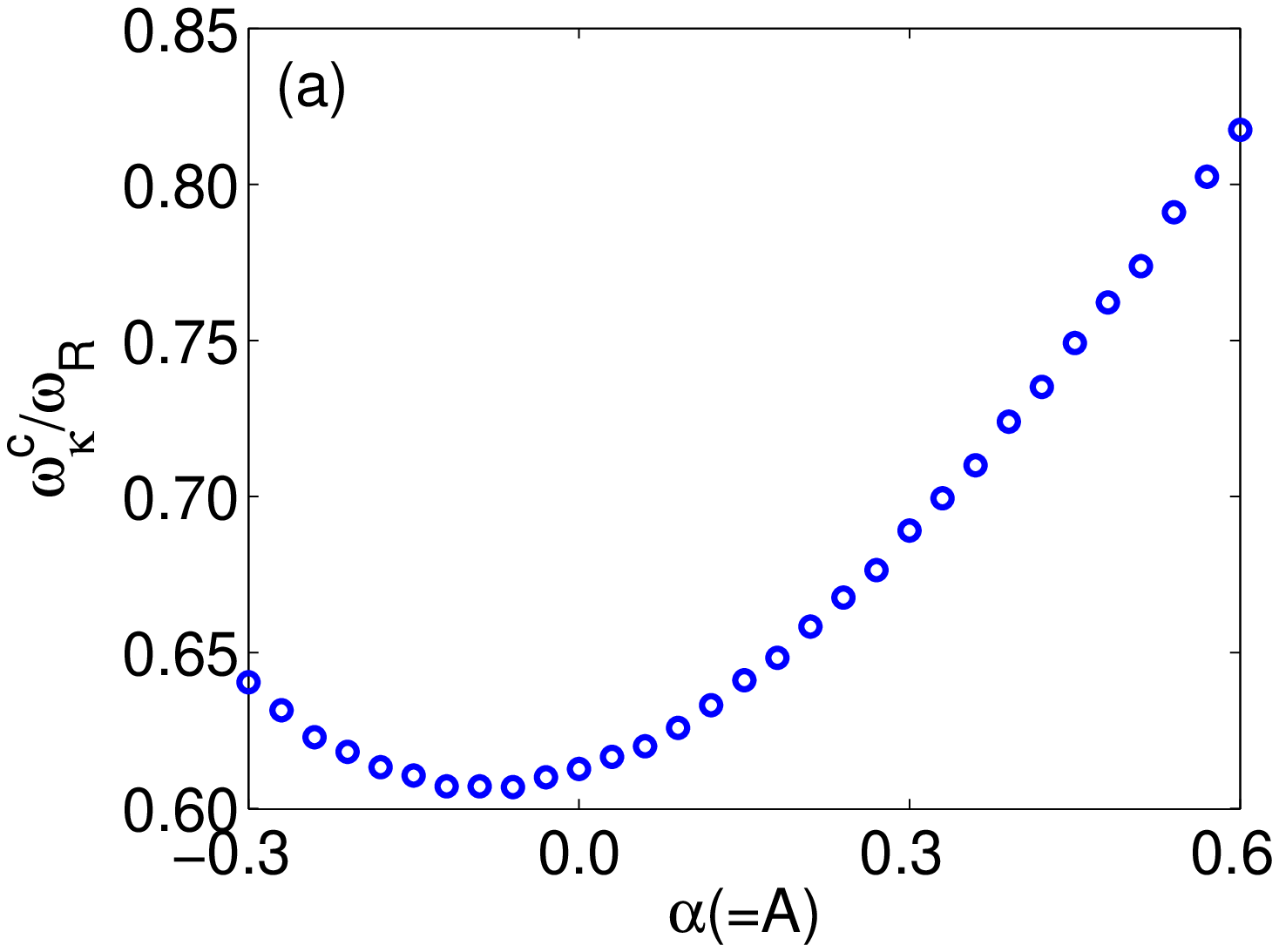}}
  \subfigure{\label{Fig2b}\includegraphics[width=0.23\textwidth]{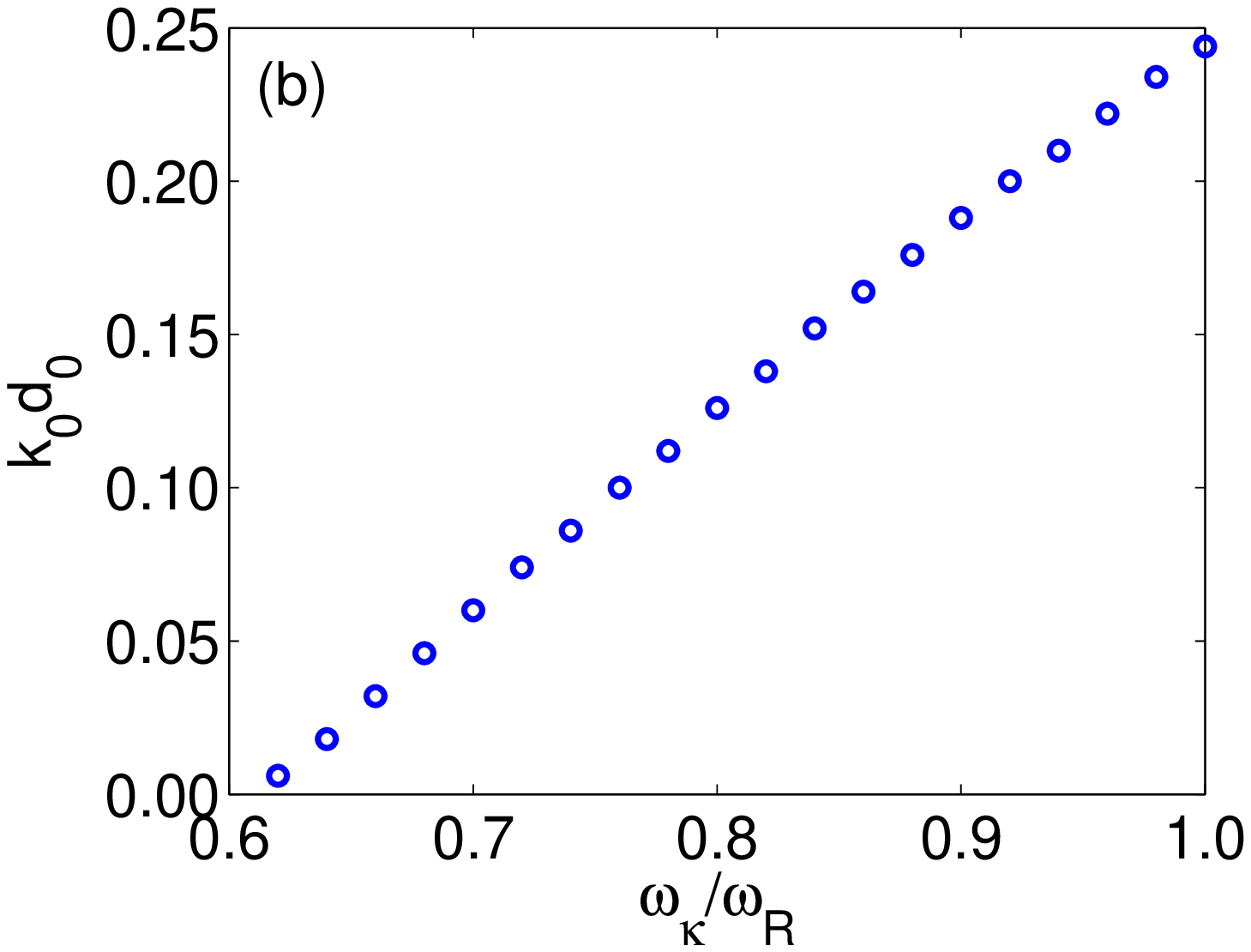}}
\caption{(Color online) (a) The critical value of chirality $\omega_\kappa^c$
versus $\alpha(=A)$, for two identical CMM plates. (b) The equilibrium distance $k_0d_0$ versus
$\omega_\kappa$ for $\alpha=A=10^{-3}$. For
$\omega_\kappa>\omega_\kappa^c=0.612\omega_R$, the
value of the equilibrium distance $k_0d_0$ corresponds to the
minimum of the energy as shown by the open circle curve in Fig.
\ref{Fig1}.
}
 \label{Fig2}
\end{figure}

Here, we consider first a special setup with two identical chiral metamaterial plates with the following parameters:
$\epsilon_1=\epsilon_2=\epsilon; \mu_1=\mu_2=\mu;
\kappa_1=\kappa_2=\kappa$. We suspect that the chirality
coefficient, $\kappa$, may provide sufficient new freedom to drive the
force to negative values (i.e., repulsive) at least for some range of
values of \textit{d}. From Eq. (\ref{energy}) it follows Ref. \cite{Rosa} that a
negative value of the Casimir force is favored by making the
quantity $I\equiv \textrm{Tr}[\mathbf{D}^{-1}(1-\mathbf{D})]$ as
negative as possible over as broad a range as possible of the parameters and the
integration variables. This quantity, \textit{I}, has the
same sign as the quantity \textit{F} given below:
\begin{equation}\label{F}
F=\frac{(r_{ss}^2+r_{pp}^2-2r_{sp}^2)e^{-2Kd}-2(r_{sp}^2+r_{ss}r_{pp})^2e^{-4Kd}}{1-(r_{ss}^2+r_{pp}^2-2r_{sp}^2)e^{-2Kd}+(r_{sp}^2+r_{ss}r_{pp})^2e^{-4Kd}}.
\end{equation}
Because $r_{sp}$ is purely real as shown in Eq.
(\ref{elements}), it is clear from Eq. (\ref{F}) that the chirality
by introducing the off-diagonal quantity $r_{sp}$ provides the
possibility, for large enough $r_{sp}$, to make the numerator in Eq.
(\ref{F}) negative, while keeping the denominator
positive. Thus, the chirality, if strong enough, is expected to lead
to a repulsive Casimir force. This expectation is confirmed by the
numerical evaluation of the interaction energy per unit area as
shown in Fig. \ref{Fig1}. Indeed, for large enough chirality
parameter, $\omega_{\kappa 1}=\omega_{\kappa
2}=0.7\omega_R$, we have a very interesting situation of an
attractive force in the range $d>d_0$ (where in the present case
$d_0=0.0586c/\omega_R$) and a repulsive case for
$d<d_0$. Thus, a stable equilibrium distance emerges, $d=d_0$,
reminiscent of the bond length in a diatomic molecule. There is a
critical value of $\omega_\kappa$, $\omega_\kappa=\omega_\kappa^c$,
such that for $\omega_\kappa<\omega_\kappa^c$ there is no repulsive
regime for any value of $d$, while for
$\omega_\kappa>\omega_\kappa^c$, there is a distance $d_0$, a
function of $\omega_\kappa$, $d_0(\omega_\kappa)$, such for
$d<d_0(\omega_\kappa)$ the force is repulsive. For the numerical
values used in our present case, the critical value
of $\omega_\kappa^c$ is equal to
$\omega_\kappa^c=0.612\omega_R$ for $\alpha(=A)=0$. As shown in Fig. \ref{Fig2a}, the critical value $\omega_\kappa^c$ is a function of the $\alpha$ with its minimum value $\omega_\kappa^c=0.607$ obtained for $\alpha\simeq-0.09$. Furthermore, the
relation $d_0$ versus $\omega_\kappa$ (for
$\omega_\kappa>\omega_\kappa^c$) is an increasing almost linear
function of $\omega_\kappa$, as shown in Fig. \ref{Fig2b}.

The question raised by the present novel approach to a possible
repulsive Casimir force is whether real chiral metamaterials can be
fabricated with $\omega_\kappa$ larger than the critical one
$\omega_\kappa^c$. Our own chiral metamaterial presented in Ref.
\cite{chiral} has an $\omega_\kappa\simeq0.3\omega_\kappa^c$.
However, this metamaterial was designed and fabricated before the
critical importance of chirality for stable Casimir nanolevitation
was even suspected; thus, there is room for
new designs to raise the value of $\omega_\kappa$ possibly
above the critical value.  We are currently working on this theme. We don't know whether or not general physical considerations restrict the size of the chirality factor $\omega_\kappa$ and thus we cannot be sure whether the critical value of $\omega_\kappa^c$ is reachable. Models based on a single loop (see the books of Lindell et al. \cite{Lakhtakia} and Serdyukov et al. \cite{Serdyukov}) produce a relation between the electric, $\alpha_{ee}$, the magnetic, $\alpha_{mm}$, and the cross polarizabilities, $\alpha_{em}$, $\alpha_{me}$: $\alpha_{ee}\alpha_{mm}=\alpha_{em}\alpha_{me}$. This relation, valid when $\omega_m=\omega_R=\omega_{\kappa R},~\gamma_m=\gamma_R=\gamma_\kappa$ and $\omega_{pl}=\alpha=0,~A\neq0$, shows that the critical value $\omega_\kappa^c$ is almost reachable under the optimum condition $A\to\omega_e^2/\omega_R^2$.

\begin{figure}[htb!]
 \centerline{\includegraphics[angle=0, width=8.cm]{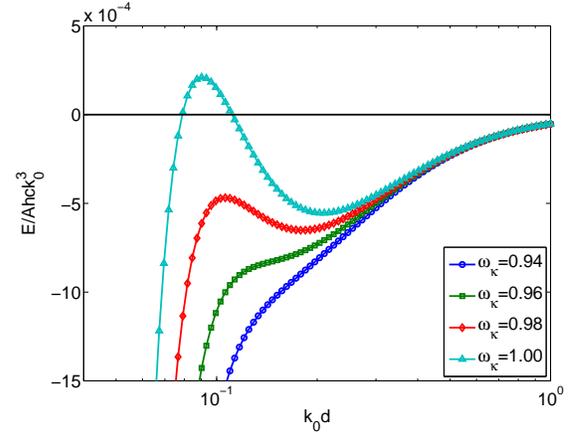}}
\caption{(Color online) Casimir interaction energy per unit area E/A
(in units of $hck_0^3$) versus  $k_0d$ of the
two identical CMM plates configuration for different chiral
strengths $\omega_\kappa$'s.
$\alpha=0,~A=0.2,~\omega_{pl}=0,~\omega_{m}=\omega_{\kappa
R}=\omega_R,~\gamma_e=\gamma_m=\gamma_\kappa=0.05\omega_R,~\omega_e=\omega_R.$}
 \label{Fig3}
\end{figure}

In Fig. \ref{Fig3} we present results for the energy per unit area
$E/A$ versus the dimensionless distance $k_0d$ for chiral
metamaterials with $\mu(\omega)$ given by Eq. (\ref{mu}) with
$\alpha=0$ and $A=0.2$. We repeat here this choice violates the
physical requirement of $\mu(\omega)\to 1$ as $\omega\to \infty$.
Nevertheless, we present these results here in order to show that unphysical frequency dependence of the response functions may produce the resulting behavior which is qualitatively different from
that presented in Fig. \ref{Fig1} in the sense that now two equilibrium
points, $d_1$ and $d_2$ ($d_1<d_2$), may appear, the first is
unstable equilibrium and the second is stable equilibrium.
Furthermore, one cannot exclude the possibility that a more
complicated $\mu(\omega)$ satisfying the condition $\mu(\infty)=1$
and producing results as those in Fig. \ref{Fig3} may exist. In spite of this unphysical
behavior of $\mu(\infty)$ ($\mu(\infty)=0.8$ instead of $\mu(\infty)=1$), one expects to produce no repulsive force if regular metamaterials (with no chirality) are employed. The reason is
that both of the interacting plates are mainly nonmagnetic with
$\mu(i\xi)<1<\epsilon(i\xi)$ at all frequencies. For a not so large
chirality ($\omega_\kappa=0.94$ (circles)), one can
easily see from Fig. \ref{Fig3} there is only an attractive Casimir
force for all distances. However, as chirality increases,
($\omega_\kappa=0.96$ (squares)), the energy tries
to develop a minimum, and the Casimir force corresponding to the
slope of the energy becomes smaller. At
$\omega_\kappa=0.98$ (diamonds), there is a minimum
at $k_0d_2\simeq 0.21$ and a maximum at
$k_0d_1\simeq 0.09$; the peak value of the energy is
less than zero, the energy value at $k_0d=\infty$. If chirality
increases further, $\omega_\kappa=1.00$
(triangles), the sign of the energy is reversed and becomes positive
in a certain range. This is an interesting case that gives a
repulsive Casimir force within a range of distances between $d_1$
and $d_2$. It forms a potential barrier to block the two interacting
plates sticking to each other. Similar results to those in Fig.
\ref{Fig3} were also obtained for the case where we used
$\epsilon(\omega)=2-\omega_e^2/(\omega^2-\omega_e^2+i\gamma\omega)$.
This frequency-dependence is obtained experimentally \cite{NIMs} for
realistic metamaterials, but only close to the resonance behavior;
such a dependence extended to $\omega\to\infty$ violates the
condition of $\epsilon(\omega)\to 1$ as $\omega\to\infty$.

In discussing these results we must keep in mind that for
$k_0d\ll1$ the main contribution to the integral in Eq. (\ref{energy}) comes from large $\xi$ and $k_\parallel$ values with the ratio $k_\parallel/\xi\gg1$, as argued by Landau et al. \cite{Landau} and confirmed by our numerical calculations. Under these conditions $k\simeq k_\parallel$ and the integrand in Eq. (\ref{energy}) takes the form $f(\xi,e^{-2k_\parallel d})$. By setting $x=2k_\parallel d$, it follows immediately from Eq. (\ref{energy}) that $E(d)/A\propto d^{-2}$ and $F(d)/A\propto d^{-3}$; the contribution of the chiral term to $f(\xi,e^{-x})$ is negative and, thus, for large enough chirality the force in the $d\to0$ limit becomes repulsive. On the other hand, in the opposite limit $d\to\infty$, because of the factor $e^{-2Kd}$, the main contribution to the integral comes from the range $0\le\xi\lesssim(c/d)$ and $0 \lesssim k_\parallel \lesssim (d^{-1})$, where the integrand tends to a constant corresponding to the $\xi=0$ values of $\epsilon(0)>1,~\mu(0)\simeq1$, and $\kappa(0)=0$. Thus in this $d\to\infty$ limit $E(d)/A\propto d^{-3}$ and $F(d)/A\propto d^{-4}$ and the force is always repulsive, since essentially only $\epsilon(0)$ matters. This analysis shows that it is crucial to employ the correct limiting values of $\epsilon(i\xi),~\mu(i\xi),~\kappa(i\xi)$ as $\xi\to\infty$ and $\xi\to0$, since these values determine the behavior of $E(d)/A$ in the limit $d\to0$ and $d\to\infty$ respectively.

In this work we have extended the Lifshitz theory to calculate
the Casimir force by including chirality terms for the first time.
We have shown that the chirality, if strong enough, is of critical
importance in producing nanolevitations under realistic frequency-dependence and correct limiting values of $\epsilon(\omega)$ and
$\mu(\omega)$. Note, the previous calculations claiming
repulsive Casimir force between metamaterials separated by vacuum
have been achieved at the expense of nonrealistic frequency-dependence and/or limiting values of $\epsilon(\omega)$ and
$\mu(\omega)$. Thus, chiral metamaterials might possibly be the main
candidates to achieve experimentally the goal of Casimir repulsion,
which  might open up many opportunities for application.


Work at Ames Laboratory was supported by the Department of Energy
(Basic Energy Sciences) under contract No.~DE-AC02-07CH11358. This
work was partially supported by the European Community FET project
PHOME (contract No.~213390), US Department of Commerce NIST
70NANB7H6138 and the US Air Force grants. The author Rongkuo Zhao
specially acknowledges the China Scholarship Council (CSC).

\bibliographystyle{apsrev}

\begin{thebibliography}{99}
\bibitem{another}
A conceptually (but not calculationally) simple way to obtain Eqs. (\ref{energy}) and (\ref{matrix}) is to calculate the surface electromagnetic eigenmodes $\omega(\mathbf{k}_\parallel)$ of this system and then to evaluate their zero point energy, $\sum_i\int d\mathbf{k}_\parallel \frac{1}{2}\hbar\omega_i(\mathbf{k}_\parallel)$.
\end{thebibliography}

\end{document}